# Inter-Rater: Software for analysis of inter-rater reliability by permutating pairs of multiple users

D.J. Arenas

*Perelman School of Medicine, University of Pennsylvania, Philadelphia, PA 19104, USA.*

Inter-Rater quantifies the reliability between multiple raters who evaluate a group of subjects. It calculates the group quantity, Fleiss kappa, and it improves on existing software by keeping information about each user and quantifying how each user agreed with the rest of the group. This is accomplished through permutations of user pairs. The software was written in Python, can be run in Linux, and the code is deposited in Zenodo and GitHub. This software can be used for evaluation of inter-rater reliability in systematic reviews, medical diagnosis algorithms, education applications, and others.

## 1. INTRODUCTION

There are many research fields and applications where subjects must be assigned to different categories (i.e. "benign" or "malignant"; "relevant" or "irrelevant"). Commonly, these assignments are performed by raters who follow objective guidelines and/or algorithms. However, sometimes the complexity of the problem makes it impossible to eliminate all subjectivity in the evaluation. This scenario applies to challenging cases of medical diagnoses, where the complexity is sufficiently high for different raters to assign different diagnoses to the same subject. Such complexity, and its resulting subjectivity, is also a challenge in psychiatric evaluations, and the reading of images for diagnostic purposes in pathology, surgical reviews, and oncology.[1-3]

This subjectivity usually requires that multiple raters evaluate the same subject, in the hopes that the group maximizes the accuracy of the assignment. Evaluation by multiple raters may also be useful in validating the clarity and objectivity of the algorithm or thought process. It is reasonable to expect higher agreement between raters for clearer and more objective guidelines. For example, one expects that if the diagnosis guidelines for hypertension are clear then several physicians would have high percentage agreement in diagnosing the same patient. Therefore, there is clearly a need for quantifying agreement between raters. And this need is not exclusive to medicine, for the necessity is also prevalent in fields such as education, where it is useful in grading.[4]

Certainly, higher percentage agreement between raters is desirable, but percentage-agreement alone is not sufficient to evaluate if the raters are using similar guidelines. As an example, consider two teachers grading student essays into two categories: "pass" or "fail", and the two teachers were instructed to use different criteria. Even if one teacher focuses solely on grammar, and the other solely on intellectual content, it would still be possible for the teachers to agree a high percentage of the time; if both teachers were to pass ~ 90% of their students, they would agree around 82% of the time even though they were following different rules. (81% of the time agreeing on "pass" and 1% agreeing on "fail").

In other words, one must consider the probability the raters agree by chance ($p_e$). Then we must assign a statistic (with a confidence interval) to the agreement. Then, we must ensure that this agreement was significantly different from randomness. One of the most common methods to achieve this goal is Cohen's kappa ($\kappa_C$). This kappa is designed for two users; it compares the actual agreement ($p_o$) between the raters from agreement by chance, and divides it over the possible improvement over chance:

$$\kappa_C = \frac{p_0 - p_e}{1 - p_e}$$

In the previous two-teacher example, if the teachers had agreed 85% of the time, then the Cohen's kappa would have been 0.22, and the confidence interval (and conclusions of significance) would depend on the number of subjects they rated. For Cohen's kappa, any value significantly above zero shows agreement between raters that was better than chance; and the closer the kappa is to one the better the agreement.

Many of the aforementioned applications use multiple raters, and therefore require an improvement over Cohen's two-rater kappa. The most commonly used quantity is the Fleiss kappa ($\kappa_F$),[5,6] a group statistic that also calculates the agreement expected from chance, and the average agreement between the multiple raters. For more information on the calculation of this quantity, refer to Appendix A.

In summary, both the Cohen kappa and Fleiss kappa are useful in ensuring that agreement between raters is significantly better than chance. Unfortunately, besides this ability, the utility of these kappas is limited. Intuitively, a value close to 1 is better, but the meaning of intermediate values (say 0.2 or 0.4) is not clear. There have been suggestions for scales ("poor", "moderate", etc.) of agreement,[7] but other authors have argued that these scales were arbitrary and unhelpful.[8] In multi-rater analysis, another shortcoming of the Fleiss-kappa is loss of information. In Fleiss analysis, the specific ratings from each user are lost in the calculation (see appendix A for the formulas); therefore potential information about one user's, or subgroups, large disagreements from the rest of the group is lost. This is unfortunate, since one application of inter-rater reliability may be evaluation of the raters of the algorithms. To our knowledge, all inter-rater reliability software packages in *R* (and other open source software), as well as professional software, do not take advantage of user specific information since they only use the Fleiss kappa in multi-raters assignment evaluations. Also, to our knowledge, these software options do not plot the results in an easy to understand fashion.

Here we present software that not only calculates the overall inter-rater reliability of the group (Fleiss' kappa), but also keeps information about each user by permutating the inter-rater reliability between one user and the others. Although it is highly unlikely that this is the first time that permutations of pairs are suggested, permutated-kappas are underutilized in the literature. In this software, a permutated-kappa is given to each user, and gives information on users who may significantly disagree from the rest of the group.

Lastly, Appendix A in the supplementary documents gives an overview of the theory behind the inter-rater reliability. This may be useful to readers for two reasons: 1) The theory behind the variance of Fleiss kappa went through several iterations,[9] and the document summarizes the most recent and accurate one. 2) The terminology from the original papers is updated and condensed into one.

## 2. IMPLEMENTATION AND ARCHITECTURE

### 2.1. Data file and categories file

We developed Inter-Rater to analyze the ratings of *N* subjects by n users. The program analyses an input data file with the *N*x*n* matrix where each row is a subject and each column corresponds to the ratings of one rater. The program automatically works for any number of raters, categories, and subjects present. Two example data files are included in the program. These files contain relevance ratings from recently submitted systematic reviews, and they are labelled as *data_SR1.txt*, and *data_SR2.txt*.

Another input file contains the possible categories for ratings (i.e. "benign", "malignant"; "1", "0", "-1"). The ratings are specified by the user, and not automatically searched by the software, so that incorrect entries in the ratings are ignored and labelled as abstinences. An example file is also provided: *categories.txt*.

It is important to note that it is not necessary that all raters evaluate every study. In the permutation of user-pairs, the program only considers subjects that were rated by the pair. Therefore it builds an $N_2$x2 matrix, where $N_2$ is only the subjects for which both raters entered one of the possible ratings. The flexibility to allow users to only rate subsets of the studies may be useful in systematic reviews where it is not possible for all raters to examine all manuscripts.

### 2.2. Running the program

The researcher runs the program in the Linux command terminal. An example of the simplest way to run the code is:

```
$ ./Inter_rater.py  -dfile data_SR1.txt
```

Various inputs are labelled with flag. The most important, and necessary flag *"-dfile"*, is the name of the data file. The user can also input a file with the possible categories of the ratings using the *"-cfile"* flag. For example:

$ ./Inter_rater.py  -dfile data_SR1.txt -cfile categories.txt

Table 1 lists all the possible flags for the program.

### 2.3. Output to screen

The program first outputs information about the entire group: The probability that each category was chosen, the average agreement between users, and the Fleiss kappa (with its confidence interval). Second, the program outputs information on each user: Her/his probability to choose each category, the Cohen kappa (and its confidence interval) for each pair permutation, and the average (and its confidence interval) of these permutations.

### 2.4. Output to graph

For ease of interpretation, the aforementioned results are also plotted to an output *jpg* file. The researcher has the option to specify the desired *y*-axis parameters from the command line (See Table 1). The user also has the option of showing the confidence interval for each permutated-kappa, and of highlighting a particular pair of raters (See Table 1).

### 2.5. Permutated-kappa tensor

An important intermediate result is a permutated-kappa tensor, *PK_tensor*[$x$][$y$][$stat$], with size $n \times n \times 2$. This tensor is such that its [$x$][$y$][0] denotes the $x,y$ pair of users, and the final index holds information about the average of the Cohen kappa. Similarly, [$x$][$y$][1] is the standard error of the kappa between users $x$ and $y$. For ease of interpretation, the aforementioned results are also plotted to an output *jpg* file. The researcher has the option to specify the desired *y*-axis parameters from the command line (See Table 1). The user also has the option of showing the confidence interval for each permutated-kappa, and of highlighting a particular pair of raters (See Table 1).

### 2.6. Python code architecture

A library file (*inter_rater_library.py*) contains all of the used functions. The functions are broken down into subgroups of calculating the Fleiss kappa and its variance, calculating the Cohen kappa, permutating around pairs of users, and plotting the final results. Appendix A, a brief discussion of the theory, shows which important functions accomplish calculation of intermediate and final results.

### 2.7. Two systematic reviews examples

To test the whole software, we used it to analyze the inter-rater reliability in abstract-exclusion from two systematic reviews. In each systematic review, the raters had to read a group of abstracts and decide whether each was relevant to their intended scope. Quick and convenient rating of the abstracts was possible using the Abstrackr website.[10] Figure 1 shows the results for analysis of the ratings from one systematic review, where four raters evaluated 1620 abstracts and decided whether each manuscript discussed food insecurity and its effect on health outcomes.[11] For each user (#0,#1,#2,...) Inter-Rater calculated the Cohen kappa between this user and the other three. Each permutation of pairs (i.e. *"(0,1)"*) are shown as a blue dot. For each user, the permutated-kappas are averaged; and these averages, along with their confidence interval, are shown in orange. The confidence interval for the group's Fleiss kappa is shown in green dashed lines.

For this set of ratings, we see that all permutations had kappas with confidence intervals higher than one, which generates confirmation of agreement above chance. Furthermore, the program shows that user *#1* had the most overall agreement with the other 3, while user *#3* had the worst. The confidence interval of user *#3*, however, was still overlapping with the others so there is no significance in his/her lower kappa. The confidence interval of the group kappa also shows that the group's overall agreement is better than chance.

Figure 2 shows the results for another systematic review of 81 abstracts by 4 raters. The results show that this group also agreed better than chance. It also shows that user *#2* was the one with most disagreement over the others. For this review, however, the permutated-kappas between users *#1* and *#2* were significantly different. This type of significant disagreement may be useful in generating discussions of the review process afterwards; for example, user *#2* may have a different training background than his/her teammates.

### 2.8. Discussion on speed: Size of the data file and number of users

The running time of Inter-Rater is proportional to $N \times n \times 2$, where $N$ is the number of subjects and $n$ is the number of raters. Using the example data files, 4 users for ~1000 subjects, the running time of Inter-Rater in a standard laptop or desktop is in the order of seconds. Most applications by human researchers (i.e. systematic reviews and medical diagnosis) will naturally have a limited amount of users and subjects; therefore, for these applications running time should not be an issue.

### 2.9. Quality control

The program was written into several functions, each of which was individually tested for correctness. The researcher can use the example data files, and the examples to run the code, to ensure that the program is running well. Figure 1 should be duplicated using *data_SR1.txt*, and Figure 2 using *data_SR2.txt*.

## 3. AVAILABILITY

**Operating system**
Linux. Ubuntu 17.10

**Programming language**
Python 2.7

**Additional system requirements**
None

**Dependencies**
numpy (1.12.1)
matplotlib (2.0.0)

**List of contributors**
All contributors are authors in the manuscript.

### 3.1. Software location

**Archive**
Name: Zenodo
Persistent identifier: https://zenodo.org/record/1227660
Licence: GNU General Public License (v3)
Publisher: Daniel J. Arenas
Version published: 1.0
Date published: 24/04/2018

**Code repository**
Name: GitHub
Identifier: https://github.com/djarenas/Inter-Rater
Licence: GNU General Public License (v3)
Date published: 24/04/2018

**Language**
English

## 4. REUSE POTENTIAL

As previously demonstrated, Inter-Rater can be used by authors of systematic review where researchers with different expertise rate several abstracts. Furthermore, the software can be used in medical-diagnosis studies where different experts come together - in fact, there are already published studies in the literature that would have benefited from the permutated-kappa analysis and this software.[1] The software can also be used by educators who want to validate algorithms for grading.

There is also the possibility of using Inter-Rater for fundamental research: for example, the group Fleiss kappa could be compared to user-averaging of the permutated-kappa. For theoretical purposes, this program could also be modified and extended into permutations of sub-groups.

Support for modification of the software is possible through the GitHub page at https://github.com/djarenas/Inter-Rater/issues. Users are also welcome and encouraged to email the corresponding author to request help using the program, analyzing results, as well as requests for specific modifications. Collaborations are always welcome.

## 5. DECLARATIONS

### 5.1. Acknowledgements

D.J.A would like to thank Chu Chuan Chiu, Jeff Hoskins, and Riccardo Fincato for useful discussions.

### 5.2. Funding statement

D.J.A. would like to thank the Gamble Scholarship for support. No funding was used for development of this program.

### 5.3. Competing interests

The authors declare they have no competing interests.

| Flag | Purpose | Example | Default |
|---|---|---|---|
| -dfile | Data filename | "data_SR1.txt" | Mandatory |
| -cfile | Categories filename | "categories.txt" | "categories.txt" |
| -ofile | Desired filename for the final jpg graph | "output_SR1.jpg" | "output_graph.jpg" |
| -ymin | Y-axis minimum for final jpg graph | 0 | 0 |
| -ymax | Y-axis maximum for final jpg graph | 1 | 1 |
| -highlight | Highlights a pair of raters and plots it black | 2,3 | "none" |
| -indbars | Individual error bars for every pair kappa | "yes" | "no" |

**Table 1.** Possible flags for running the program, examples, and their default.

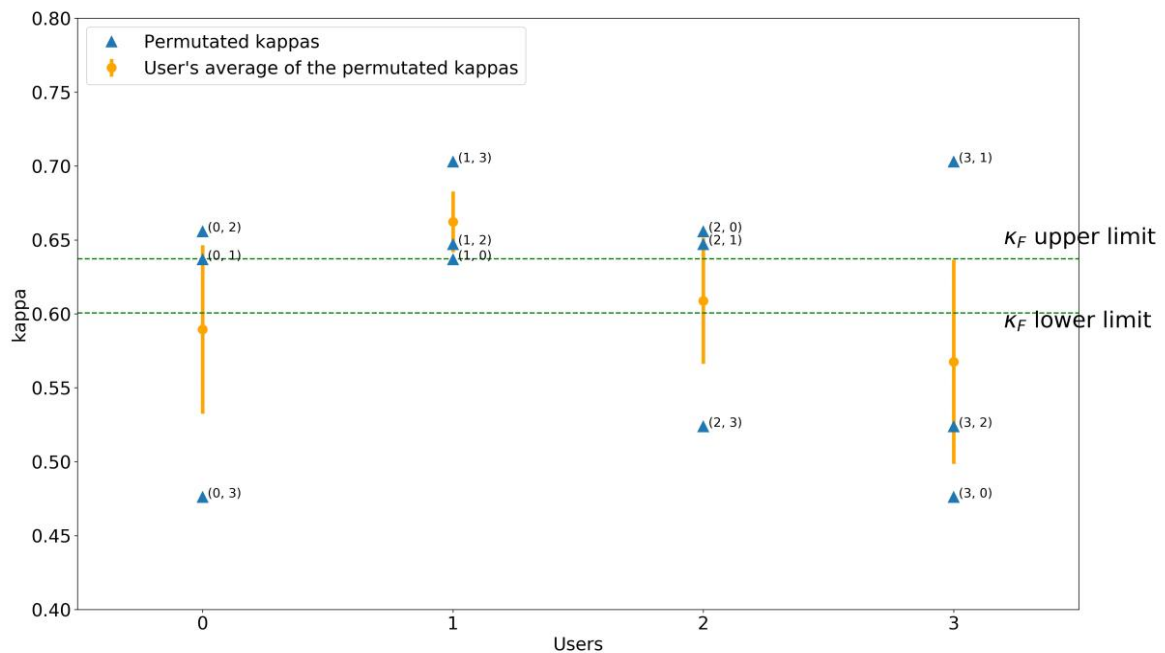

**Figure 1.** Evaluation of multi-user inter-rater reliability for four authors in a systematic review of 1620 abstracts. Permutated-kappas are shown for each user-combination as blue triangles. The average, and 95% CI, of these permutations are shown in orange. The 95% CI of the group Fleiss kappa are shown in dashed green lines.

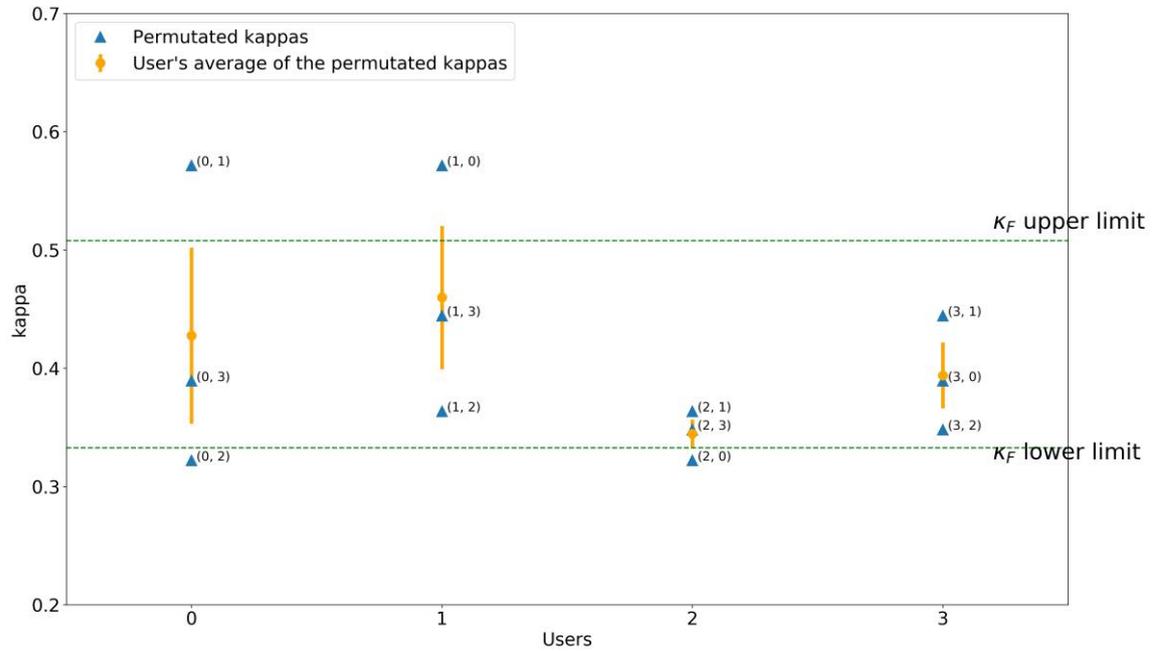

**Figure 2.** Evaluation of multi-user inter-rater reliability for four authors in a systematic review of 81 manuscripts. Permutated kappas are shown for each user combination as blue points. The average, and 95% CI, of these permutations are shown in orange. The 95% CI of the group Fleiss kappa are shown in dashed green lines.

# Appendix A: Background

## Cohen's kappa ($\kappa_c$)
*Python function: "Calculate_Cohen_kappa".*

Cohen's kappa metric ($\kappa_c$) is used for two raters evaluating $N$ subjects. They assign each subject to one of $k$ categories (i.e. "good", "bad", …). The kappa metric gives information on whether the users agreed based on random assignments or based on a common criterion. One must first calculate the probability of random agreement by counting the number of times that user #1 (or #2) choose category $j$, ($n_{j1,2}$), and convert it to a probability:

$$p_{j1,2} = \frac{n_{j1,2}}{N} \quad [1]$$

The probability that the two users agreed by chance, $p_e$, is then:

$$p_e = \sum_{j=1}^{k} p_{j,1} p_{j,2}. \quad [2]$$

One then counts the number of subjects for which the users agreed, and calculate the probability of agreement, $p_o$. We divide how much higher the agreement was compared to random agreement, $p_o - p_e$, over the possible improvement over random agreement ($1 - p_e$).

$$\kappa_c = \frac{p_o - p_e}{1 - p_e}. \quad [3]$$

A value of zero therefore represents no improvement over random assignments, and a value of 1 indicates perfect agreement. It should be specified that this metric is designed for categorical ratings (i.e. "north", "south", "east", "west"), that is, ratings that are not part of a scale. An example of a scale would be ("hot", "medium", "cold").

## Fleiss's kappa ($\kappa_F$)
*Python program: "Calculate_Fleiss_kappa.py"*

Fleiss kappa (denoted here as $\kappa_f$), similarly to Cohen's kappa, is a metric designed for categorical ratings. Unlike Cohen's, it can be extended from two raters to multiple $n$ raters. It also gives the option that not all raters have to grade every subject. Similarly to Cohen kappa, it ranges from negative infinity to one.

To calculate Fleiss kappa, we first build a matrix where we count the number of ratings for each category for each abstract. The resulting $n_{ij}$ matrix is such that row $i$ denotes study $i$, and column $j$ denotes category $j$. Table A1 shows an example for ratings from a systematic review, where raters evaluate a study's relevance. Five studies, $N = 5$, were studied by four raters ($n = 4$), into three categories ($k = 3$); the resulting $n_{ij}$ matrix is shown.

|  | User 1 | User 2 | User 3 | User 4 |  | $j = 1$ |  | $j=2$ |  |
|---|---|---|---|---|---|---|---|---|---|
| Study 1 | "yes" | "maybe" | "no" | "no" | "---->" | 1 | 1 | 2 |  |
| Study 2 | "yes" | "yes" | "yes" | "yes" | "---->" | 4 | 0 | 0 |  |
| Study 3 | "no" | "maybe" | "no" | "no" | "---->" | 0 | 1 | 3 |  |
| Study 4 | "no" | "yes" | "no" | "yes" | "---->" | 2 | 0 | 2 |  |
| Study 5 | "yes" | "no" | "no" | "no" | "---->" | 1 | 0 | 3 |  |

**Table A1**. An example output where the number of ratings is counted for each category $j$ for each study $i$. This corresponds to the function "*Calculate_nijmatrix*" in the python program.

After obtaining the $n_{ij}$ matrix, we calculate the agreement for each subject, $i$, between the $n$ raters for $k$ possible categories:

$$P_i = \frac{1}{n(n-1)} \sum_{j=1}^{k} n_{ij}(n_{ij} - 1), \quad [4]$$

and average for $N$ studies:

$$P_o = \frac{1}{N} \sum_{i=1}^{N} P_i. \quad [5]$$

To calculate the probability of random assignments, one firsts calculates the probability that a rater assigns a study to category $j$. We accomplish this by finding the fraction of raters that assigned study $i$ to category $j$:

$$p_{ij} = \frac{1}{n} n_{ij}, \quad [6]$$

and estimate the probability by averaging this fraction over all studies:

$$p_j = \frac{1}{N} \sum_{i=1}^{N} p_{ij} \quad [7]$$

Table A2 uses the previous $n_{ij}$ matrix example to build the $P_i$ column and $p_j$ row

$n_{ij}$

| | | | $P_i$ |
|---|---|---|---|
| 1 | 1 | 2 | 0.166 |
| 4 | 0 | 0 | 1 |
| 0 | 1 | 3 | 0.5 |
| 2 | 0 | 2 | 0.333 |
| 1 | 0 | 3 | 0.5 |

$p_j$

| 0.4 | 0.1 | 0.5 |
|---|---|---|

**Table A2**. $P_i$ column is the agreement between raters for study $i$, and $p_j$ is the probability that category $j$ was chosen by the raters. This corresponds to the functions "*Calculate_Pi*" and "*Calculate_pj_row*" in the python program.

If the probability that a rater chooses category $j$ is $p_j$, then the probability that two raters pick category $j$ for the same study is $p_j^2$. There are $k$ possibilities or agreement and therefore the estimated probability for picking the same study randomly, $P_e$, is:

$$P_e = \sum_{j=1}^{k} p_j^2 \quad [8]$$

The Fleiss kappa value is then obtained by measuring how far above chance the raters agreed ($P_o - P_e$) over the possible improvement over random chance ($1-P_e$):

$$\kappa_f = \frac{P_o - P_e}{1 - P_e}, \quad [9]$$

## Variance of Fleiss Kappa

*Python_program: "Calculate_Fleiss_kappa_var"*

The variance of $\kappa_f$ was investigated by Fleiss et al. in his 1979 manuscript.[1] They derived the large $N$ approximation:

$$Var(\kappa_f) = \frac{2 * \lfloor \sum_j (p_j q_j)^2 - \sum_j p_j q_j (q_j - p_j) \rfloor}{Nn(n-1) \sum_j (p_j q_j)^2}, \quad [10]$$

where $q_j = 1 - p_j$. It is straightforward to show that:

$$\sum_j p_j q_j = 1 - P_e \quad [11]$$

And that equation 10 can be simplified to

$$Var(\kappa_f) = \frac{2 * \lfloor (1-P_e)^2 + 3P_e - 2\sum_j p_j^3 - 1 \rfloor}{Nn(n-1)(1-P_e)^2}, \quad [12]$$

It should be mentioned that in the literature, the derivation of the $Var(\kappa_f)$ went through several iterations where Fleiss himself noticed that some of his previous work, and his peers, had derived expressions of $Var(\kappa_F)$ that overestimated the quantity[2].

## Notes on variance for Cohen Kappa

*Python function: "Calculate_Cohen_kappa_simplisticSE" and "Calculate_Cohen_kappa_SE"*

---

[1] "Large sample variance of kappa in the case of different sets of raters.." http://psycnet.apa.org/record/1979-32706-001. Accessed 17 Apr. 2018.
[2] "Large sample variance of kappa in the case of different sets of raters.." http://psycnet.apa.org/record/1979-32706-001. Accessed 17 Apr. 2018.

Some highly cited manuscripts[3] use a simplistic estimate for the variance of Cohen's kappa. Their equation is:

$$Var(\kappa_C) = \frac{p_o(1-p_o)}{N((1-p_e)^4)}, \qquad [11]$$

Unfortunately, neither a reference nor a derivation is presented to justify this expression. Therefore we do not use this equation in our main program. We have made it available as a function and labeled as a simplistic option. In our main program we use instead the general equation [12] from Fleiss 1979 paper for two raters.[4] The simplistic, and commonly used, expression for the variance is definitely worth bringing up since this quantity overestimates the variance compared to the rigorous derivation of Fleiss et al.

---

[3] "Interrater reliability: the kappa statistic. - NCBI." https://www.ncbi.nlm.nih.gov/pubmed/23092060. Accessed 17 Apr. 2018.

[4] "Large sample variance of Kappa in cases of different sets of...." 19 Dec. 2017, https://www.researchgate.net/publication/232601920_Large_sample_variance_of_Kappa_in_cases_of_different_sets_of_raters. Accessed 17 Apr. 2018.